\documentclass[12pt]{iopart}

\usepackage{epsfig}

\begin{document}

\title[Heat Capacity and
Magnetic Phase Diagram of Y$_2$BaCuO$_5$]{Heat Capacity and
Magnetic Phase Diagram of the Low-Dimensional Antiferromagnet
Y$_2$BaCuO$_5$}

\author{W. Knafo$^{1,2,3}$, C.
Meingast$^{1}$, A. Inaba$^{4}$, Th. Wolf$^{1}$, and H. v.
L\"{o}hneysen$^{1,2}$}

\address{$^{1}$ Forschungszentrum Karlsruhe, Institut f\"{u}r Festk\"{o}rperphysik, D-76021 Karlsruhe, Germany\\
$^{2}$ Physikalisches Institut, Universit\"{a}t Karlsruhe, D-76128
Karlsruhe, Germany\\
$^{3}$ Laboratoire National des Champs Magn\'{e}tiques
Puls\'{e}s, 143 avenue de Rangueil, 31400 Toulouse, France\\
$^{4}$ Research Center for Molecular Thermodynamics, Graduate
School of Science, Osaka University, Toyonaka, Osaka 560-0043,
Japan}


\noindent{\it \today}


\begin{abstract}

A study by specific heat of a polycrystalline sample of the
low-dimensional magnetic system Y$_2$BaCuO$_5$ is presented.
Magnetic fields up to 14 T are applied and permit to extract the
($T$,$H$) phase diagram. Below $\mu_0H^*\simeq2$ T, the N\'eel
temperature, associated with a three-dimensional antiferromagnetic
long-range ordering, is constant and equals $T_N=15.6$ K. Above
$H^*$, $T_N$ increases linearly with $H$ and a field-induced
increase of the entropy at $T_N$ is related to the presence of an
isosbestic point at $T_X\simeq20$ K, where all the specific heat
curves cross. A comparison is made between Y$_2$BaCuO$_5$ and the
quasi-two-dimensional magnetic systems BaNi$_{2}$V$_{2}$O$_{8}$,
Sr$_2$CuO$_2$Cl$_2$, and Pr$_2$CuO$_4$, for which very similar
phase diagrams have been reported. An effective field-induced
magnetic anisotropy is proposed to explain these phase diagrams.

\end{abstract}

\section{Introduction}

Because of their layered structure, the undoped high-temperature
superconducting cuprates are low-dimensional magnetic systems.
Indeed, Cu-O-Cu superexchange paths within the planes are
responsible for a strong two-dimensional (2D) magnetic exchange
$J\simeq$ 1000 K between the $S=1/2$ spins of the Cu$^{2+}$ ions.
In the undoped state, three-dimensional (3D) long range ordering
occurs below a N\'{e}el temperature $T_N$ of about several
hundreds Kelvin \cite{jurgens89,nakano94}, due to a small
additional magnetic exchange $J'$ between the layers. At the
magnetic quantum phase transition of some heavy-fermion systems,
the appearance of superconductivity is believed to arise from an
enhancement of the magnetic fluctuations
\cite{thalmeier05,flouquet06}. Hence, a better understanding of
the magnetic properties of the high-$T_C$ cuprates could be of
primary importance to elucidate why superconductivity develops in
these systems \cite{storey07}. However, the magnetic energy scales
are rather high (several hundreds Kelvin) and their investigation
is difficult to perform, due to the loss of oxygen and to the
sample melting at high temperatures.

\begin{figure}[t]
    \centering
    \epsfig{file=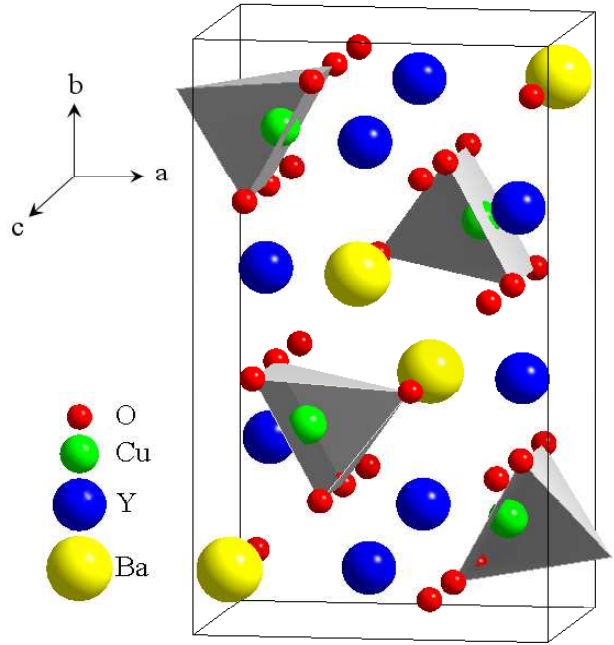,width=80mm}
    \caption{(color online) Lattice unit cell of Y$_2$BaCuO$_5$,
    where the CuO$_5$ pyramids are represented in gray.}
    \label{structure}
\end{figure}

A possible alternative is to study the magnetic properties of
systems similar to the high-$T_C$ cuprates, but with much smaller
magnetic energy scales. One of them is the "green phase" compound
Y$_2$BaCuO$_5$, which is known as an impurity phase of the
high-$T_C$ YBa$_2$Cu$_3$O$_{6+\delta}$ \cite{liang04} and whose
green color indicates its insulating character. As in the
high-$T_C$ cuprates, the magnetic properties of Y$_2$BaCuO$_5$ are
strongly low-dimensional and originate from the $S=1/2$ Cu$^{2+}$
ions. Indeed, broad anomalies in the magnetic susceptibility and
in the specific heat, whose maxima were reported at
$T_{max}\simeq30$ K \cite{ong88} and $T^\prime_{max}\simeq20$ K
\cite{goya96}, respectively, are believed to be due to the
low-dimensional magnetic exchange of Y$_2$BaCuO$_5$. However, and
contrary to the layered high $T_C$ cuprates where the magnetic
exchange is quasi-2D, Y$_2$BaCuO$_5$ has a rather complex 3D
lattice structure, shown in Fig. \ref{structure}, and the question
whether the dominant magnetic interactions are one-dimensional
(1D) or 2D is still open. Three kinds of Cu-O-O-Cu superexchange
paths, either 1D or 2D were suggested \cite{meyer90}. It is
unclear, however, which one of the three paths is dominating. The
fact that, in Y$_2$BaCuO$_5$, the superexchange paths go through
two oxygens, contrary to one oxygen for
YBa$_2$Cu$_3$O$_{6+\delta}$, explains the smaller magnetic energy
scales \cite{meyer90}. At lower temperatures, 3D antiferromagnetic
long-range ordering sets in below the N\'{e}el temperature
$T_N\simeq15.5$ K \cite{meyer90,gros92}, induced by the
combination of the strong low-dimensional and additional tiny 3D
magnetic interactions, but also of spin anisotropy, as we will
show here.

In this article, we present a study of the specific heat of
Y$_2$BaCuO$_5$ under magnetic fields up to 14 T, which permits to
extract the ($T,H$) phase diagram of this system. In the
discussion, we compare the phase diagram of Y$_2$BaCuO$_5$ to
similar phase diagrams obtained for other low-dimensional systems,
and we propose to explain them using the picture of an effective
field-induced anisotropy.

\section{Experimental details}

The polycrystalline sample of Y$_2$BaCuO$_5$ studied here was
synthesized by the direct method in air. Appropriate amounts of
Y$_2$O$_3$, BaCO$_3$, and CuO were mixed intensely, then pressed
into pellets, and finally reaction sintered with increasing
temperature steps between 750 and 960 $^\circ$C, without
intermediate grinding. X-ray powder diffractometry did not show
any trace of impurity phases. Except for 730 ppm Sr and 222 ppm
Fe, no other impurities were detected by x-ray fluorescence
analysis. The specific heat was measured under magnetic fields up
to 14 T using a Physical Properties Measurement System (PPMS) from
Quantum Design, and using an adiabatic calorimeter at zero field
up to 400 K \cite{matsuo98}. The specific heats below 200 K
obtained by both methods agree well with each other. To enhance
the resolution in the vicinity of the magnetic phase transitions,
the relaxation curves from the PPMS were analyzed following a
procedure similar to the one proposed by Lashley et al.
\cite{lashley03}. Here we will show only the PPMS data.

\section{Results}
\label{results}

\begin{figure}[t]
    \centering
    \epsfig{file=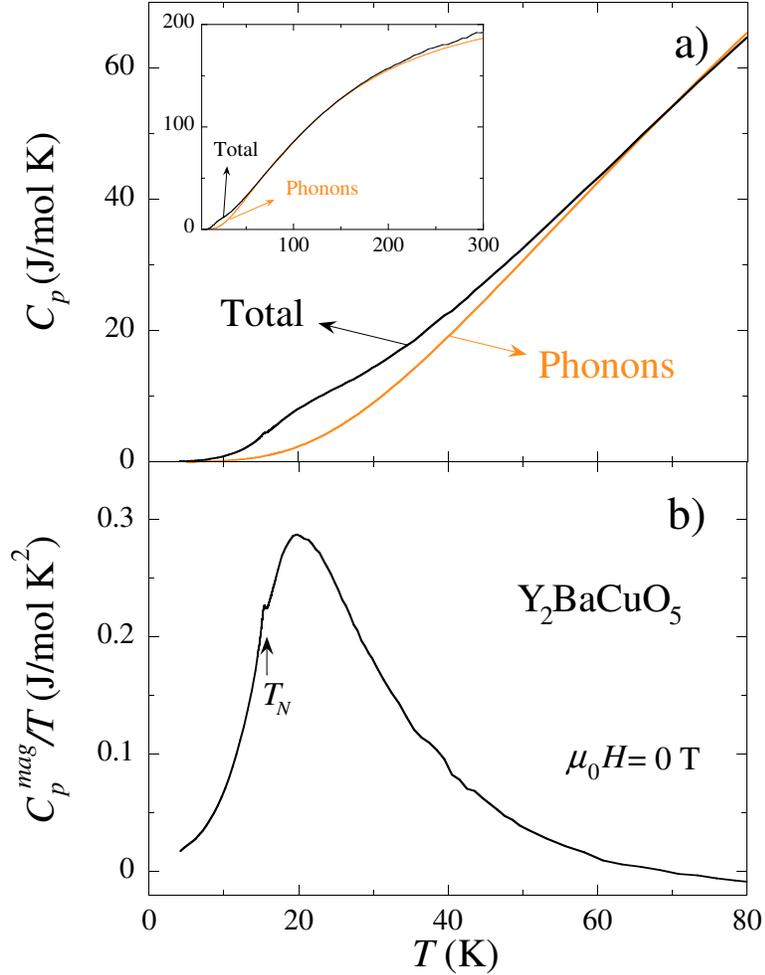,width=100mm}
    \caption{(color online) (a) Total and lattice specific heat
    of Y$_2$BaCuO$_{5}$, in a $C_p$ versus $T$ plot, with $T$ up to 80 K; data
    are plotted up to 300 K in the Inset. (b) Magnetic
    specific heat $C_p^{mag}$ of Y$_2$BaCuO$_{5}$, in a
    $C_p^{mag}/T$ versus $T$ plot. Above 60 K, $C^{mag}_p$ is less than 2 $\%$ of the total specific heat
    $C_p$.}
    \label{spec_heat_mag}
\end{figure}

\begin{figure}[t]
    \centering
    \epsfig{file=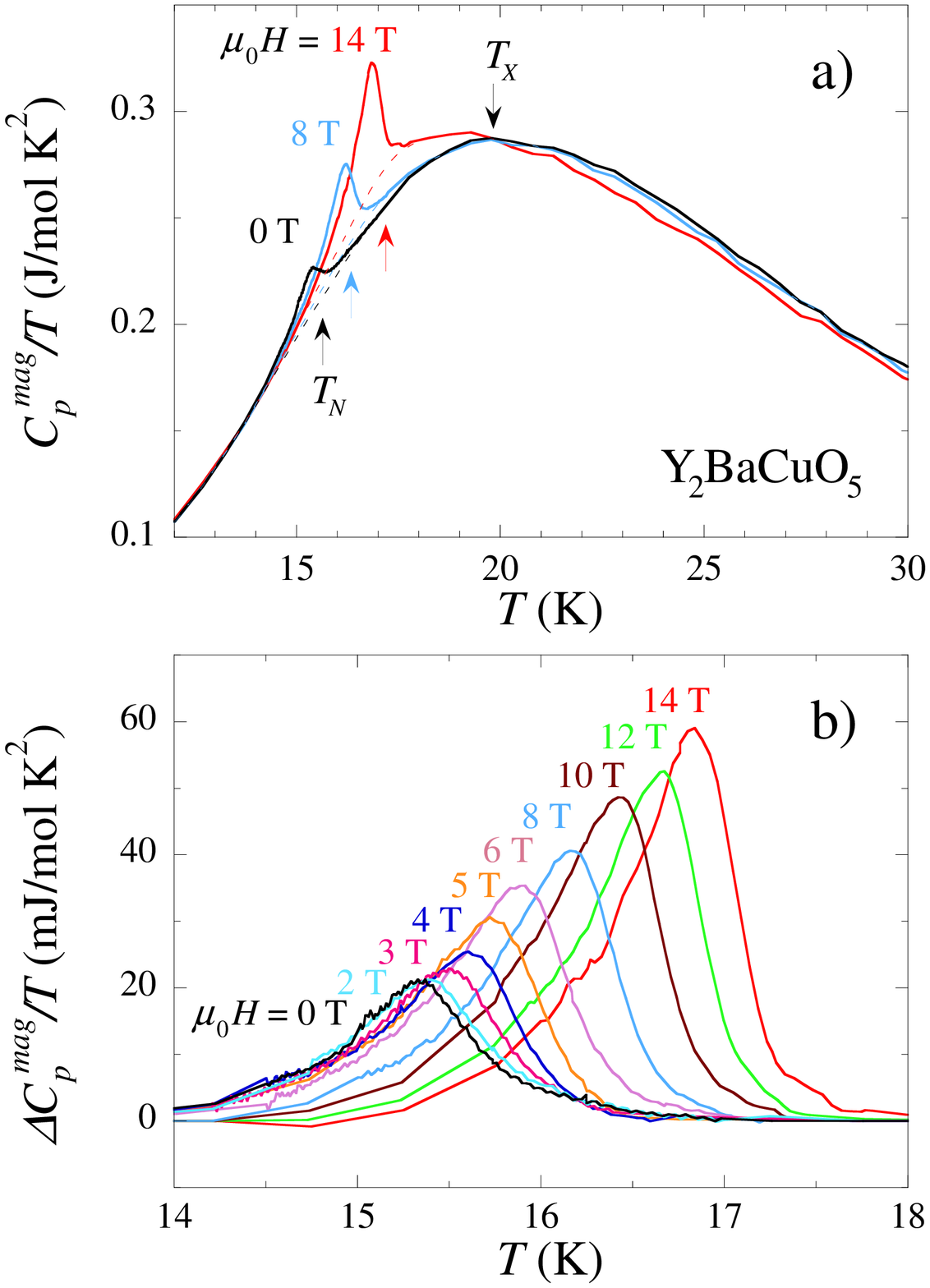,width=100mm}
    \caption{(color online) (a) Magnetic
    specific heat $C_p^{mag}$ of Y$_2$BaCuO$_{5}$, in a
    $C_p^{mag}/T$ versus $T$ plot, at $\mu_0H=0,8$, and 14 T. The
    dotted lines indicate the background used to separate the anomaly
    at $T_N$.(b) Anomaly $\Delta C_p^{mag}$ of Y$_2$BaCuO$_{5}$ at
    $T_N$, in a $\Delta C_p^{mag}/T$ versus $T$ plot, for magnetic
    fields $0\leq\mu_0H\leq14$ T.}
    \label{spec_heat_TN}
\end{figure}

Fig. \ref{spec_heat_mag} (a) shows the total specific heat
$C_p(T)$ of Y$_2$BaCuO$_5$ measured at zero magnetic field,
together with the phonon background $C^{ph}_p(T)$. $C^{ph}_p(T)$
was estimated using an appropriate scaling of the non-magnetic
specific heat of YBa$_2$Cu$_3$O$_{7}$ \cite{notephonons}. Fig.
\ref{spec_heat_mag} (a) shows $C_p(T)$ and $C^{ph}_p(T)$ for
temperatures up to 80 K, while the Inset shows the data for
temperatures up to 300 K. These plots indicate that the specific
heat of Y$_2$BaCuO$_{5}$ is almost purely phononic above 70 K and
that a magnetic signal develops below roughly 70 K. In Fig.
\ref{spec_heat_mag} (b), the magnetic contribution to the specific
heat, calculated using $C^{mag}_p(T)=C_p(T)-C^{ph}_p(T)$, is shown
in a $C^{mag}_p(T)/T$ versus $T$ plot. $C^{mag}_p$ is
characterized by a broad anomaly, whose maximum occurs at about 20
K, and by a small jump at $T_N\simeq15$ K, as typically observed
for low-dimensional systems \cite{bloembergen77,knafo07a}. The
integration of $C^{mag}_p(T)/T$ up to 70 K leads to an entropy
change of $\Delta S_{mag}\simeq6.7$ J/mol K, which equals, within
15 \%, the total entropy $R\rm{ln}2\simeq5.76$ J/mol K expected
for the $S=1/2$ spins and shows that our estimation of the phonon
background is reasonable \cite{notephonons}. Fig.
\ref{spec_heat_TN} focuses on the tiny specific-heat anomaly at
the 3D long-range antiferromagnetic ordering transition $T_N$,
which was first reported by Gros et al. \cite{gros92}. In the
following, we will concentrate on the effects of the magnetic
field on this anomaly.

In Fig. \ref{spec_heat_TN} (a), the magnetic specific heat of
Y$_2$BaCuO$_5$ is shown for the magnetic fields $\mu_0H=0$, 8, and
14 T in a $C^{mag}_p(T)/T$ versus $T$ plot. The N\'{e}el
temperature $T_N$ and the size of the anomaly at the
antiferromagnetic transition both increase with increasing $H$.
Above $T_N$, all the curves cross at an isosbestic point
\cite{vollhardt97} at $T_X\simeq20$ K, which will be discussed in
Section \ref{discussion}. For each magnetic field, an appropriate
background (dotted lines in Fig. \ref{spec_heat_TN} (a)) is used
to obtain the anomaly $\Delta C^{mag}_p(T)$ associated with the
3D magnetic ordering. In Fig. \ref{spec_heat_TN} (b), a plot of
$\Delta C^{mag}_p(T)/T$ obtained for various fields up to 14 T
emphasizes the field-induced increases of $T_{N}$ and of the size
of the anomaly at the antiferromagnetic transition. Its typical
width, of about 0.8 K (full width at half maximum), is almost
unaffected by the magnetic field and could arise from sample
imperfections (strains, impurities...) or from the polycrystalline
nature of the sample.

\begin{figure}[t]
    \centering
    \epsfig{file=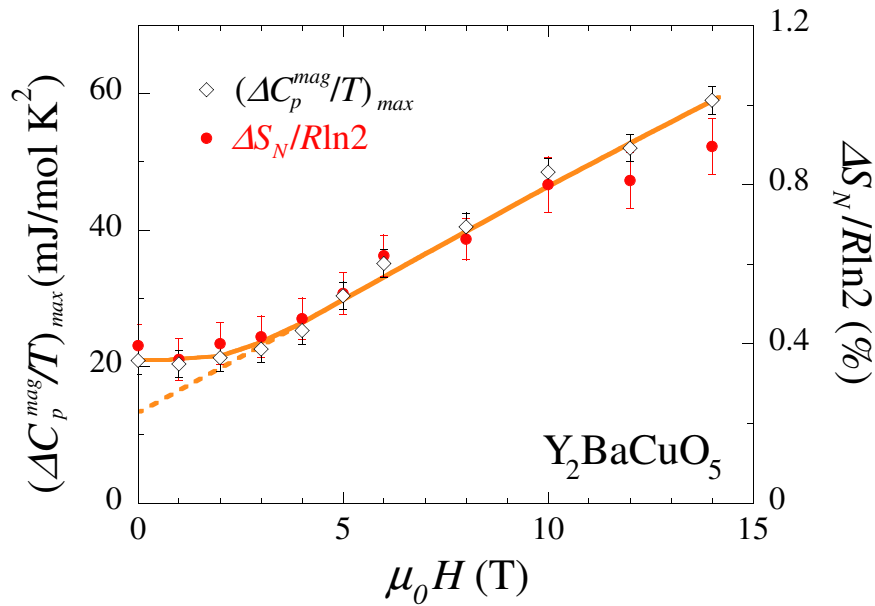,width=115mm}
    \caption{(color online) Variation with $H$ of the jump $(\Delta C_p^{mag}/T)_{max}$
    in the specific heat at the N\'{e}el ordering
    and of the associated entropy change $\Delta S_{N}$.}
    \label{entropy}
\end{figure}

In Fig. \ref{entropy}, the jump $(\Delta C_p^{mag}/T)_{max}$ in
the specific heat at $T_N$ and the associated entropy change
$\Delta S_{N}$ \cite{noteentropy} are plotted as a function of
$H$. Since the width of the anomaly is almost unaffected by $H$,
similar field dependences are obtained for $(\Delta
C_p^{mag}/T)_{max}$ and $\Delta S_{N}$: both are nearly constant
for $\mu_0H\leq \mu_0H^* \approx$ 2\,T and then increase roughly
linearly with $H$ for $H\geq H^*$. The small value $\Delta
S_N(H=0)/\Delta S_{mag}=(4.0\pm0.5)*10^{-3}$ is a consequence of
the low-dimensional character of the magnetic exchange
\cite{bloembergen77}, $\Delta S_N$ being enhanced by a factor 2.5
at $\mu_0H=14$ T.

In Fig. \ref{phase_diagram}, the $T$-$H$ phase diagram of
Y$_2$BaCuO$_5$ is shown for magnetic fields up to 14 T, $T_N$
being defined at the minimum of slope of $\Delta C_p^{mag}/T$. The
N\'{e}el temperature $T_N$ is independent of $H$ for
$\mu_0H\leq\mu_0H^*\simeq2$ T, with $T_N(H=0)=15.6\pm0.1$ K, and
increases linearly with $H$ for $H\geq H^*$, where
$T_N(H)=T_{N,0}+aH$, with $T_{N,0}=15.3\pm0.1$ K and
$a=0.14\pm0.02$ K/T (above 10 T a slight deviation from this
regime is observed). In the next Section, the field-induced
increases of $T_{N}$ and $\Delta S_{N}$ will be qualitatively
explained using the picture of a field-induced anisotropy.

\section{Discussion} \label{discussion}

\subsection{Phase diagram of Y$_2$BaCuO$_5$ - explanation in terms of field-induced anisotropy}
\label{discussionphasediagram}

\begin{figure}[t]
    \centering
    \epsfig{file=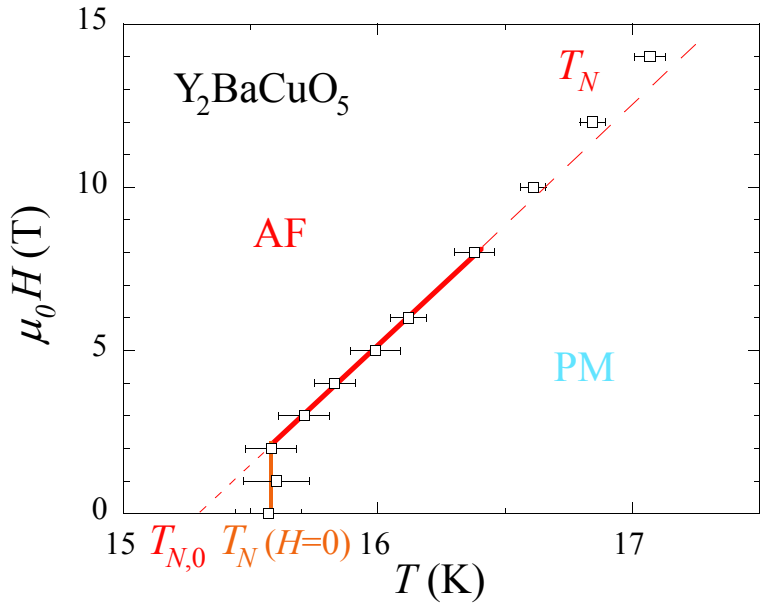,width=100mm}
    \caption{(color online) ($T,H$) phase diagram of Y$_2$BaCuO$_5$
    (polycrystal) obtained by specific-heat measurements (AF = antiferromagnetic phase,
    PM = high-temperature paramagnetic regime).}
    \label{phase_diagram}
\end{figure}

The phase diagram of the low dimensional magnetic system
Y$_2$BaCuO$_5$ (Fig. \ref{phase_diagram}) is very similar to those
of the quasi-2D magnetic systems BaNi$_{2}$V$_{2}$O$_{8}$
\cite{knafo07a}, Sr$_2$CuO$_2$Cl$_2$ \cite{suh95}, and
Pr$_2$CuO$_4$ \cite{sumarlin95} (see Section
\ref{similaritiesquasi2D}), and we interpret them similarly, using
a picture introduced 30 years ago by Villain and Loveluck for
quasi-1D antiferromagnetic systems \cite{villain77}. In this
picture, the increase of $T_N(H)$ is induced by a reduction of the
spin fluctuations parallel to $\mathbf{H}$, due to an alignment of
the antiferromagnetic fluctuations, as well as the
antiferromagnetically coupled static spins, perpendicular to
$\mathbf{H}$. This effect can be described by an effective
field-induced anisotropy, whose easy axis is $\perp\mathbf{H}$
\cite{degroot90} and which competes with the intrinsic anisotropy
of the system. As long as the field-induced anisotropy is weaker
than the intrinsic anisotropy, i.e., for $H\leq H^*$, $T_N(H)$ is
unaffected by the magnetic field and the spins align along the
intrinsic easy axes. For $H\geq H^*$, the field-induced anisotropy
is stronger than the intrinsic anisotropy and $T_N(H)$ is
controlled by $H$, the spins being aligned along the field-induced
easy axes ($\perp\mathbf{H}$).

The crossing point of the specific-heat data of Y$_2$BaCuO$_5$ at
$T_X\simeq20$ K (Fig. \ref{spec_heat_TN} (a)), is, to our
knowledge, the first isosbestic point \cite{vollhardt97} reported
in the thermodynamic properties of a low-dimensional magnetic
system. This effect is the consequence of a transfer of the
specific-heat weight, with respect to the conservation of the
magnetic entropy $\Delta S_{mag}$. Indeed, our data show that the
application of a magnetic field leads to a gain of entropy at
$T_N$, i.e. below $T_X$, which equals approximatively the loss of
entropy above $T_X$. Consistently with the picture introduced
above, we propose that the isosbestic point at $T_X$ is due to a
field-induced transfer of the specific-heat weight from the large
and broad low-dimensional short-range ordering anomaly to the
small 3D long-range ordering anomaly at $T_N$.

Little is known microscopically about the magnetic properties of
Y$_2$BaCuO$_5$, i.e., about the exact nature of the exchange
interactions and of the magnetic anisotropy. The non-layered
crystal structure of Y$_2$BaCuO$_5$ (see Fig. \ref{structure})
precludes a prediction of the superexchange paths, the dominant
paths being probably 1D or 2D \cite{meyer90}. The shapes of the
magnetic susceptibility \cite{ong88} and of the specific heat
\cite{goya96} just indicate that the dominant exchange
interactions have a low-dimensional character, either 1D or 2D
\cite{meyer90}. Some authors suggested a 2D character of the
exchange and an XY anisotropy, from appropriate fits of the
magnetic susceptibility \cite{ong88,baum04} and of the ESR
linewidth \cite{ohta95}, respectively, but these results cannot be
considered as definitive proofs. The nature of the
antiferromagnetic ordering below $T_N$ is also unclear, since
several structures have been proposed, where the spins are aligned
either in the ($\mathbf{a},\mathbf{c}$) plane
\cite{meyer90,chattopadhyay89} or along $\mathbf{c}$
\cite{golosovsky93} (cf. Ref. \cite{ohta95} for a summary of the
proposed magnetic structures). Further experimental studies on
single crystals, such as by magnetization and neutron scattering
techniques, are necessary to determine unambiguously the nature of
the spin anisotropy and of the magnetic exchange in
Y$_2$BaCuO$_5$, which can not be predicted from the complex
three-dimensional structure of Y$_2$BaCuO$_5$ (Fig.
\ref{structure}). However, we interpret the high-field increase of
$T_N(H)$ in Y$_2$BaCuO$_5$ as a consequence of a field-induced
anisotropy, a picture that works for both quasi-1D systems
\cite{villain77} and quasi-2D systems \cite{knafo07a}, the
intrinsic spin anisotropy being always ultimately of Ising-kind. A
knowledge of the nature of the intrinsic anisotropy and of the
magnetic exchange would be necessary for a more quantitative
understanding of the properties of Y$_2$BaCuO$_5$.

An additional difficulty arises from the fact that the results
presented here were obtained on a polycrystalline sample of
Y$_2$BaCuO$_5$, so that our phase diagram is equivalent to take
the average of the phase diagrams of a single crystal over all
possible field directions. In the quasi-2D magnet
BaNi$_{2}$V$_{2}$O$_{8}$, there is hardly any modification of the
magnetic properties when $\mathbf{H}\parallel\mathbf{c}$ (hard
axis) \cite{knafo07b}, which implies that a phase diagram obtained
with a polycrystalline sample would be similar to the phase
diagram reported for $\mathbf{H}\perp\mathbf{c}$ \cite{knafo07a},
possibly with a slight broadening of the transition. By analogy,
we believe that it is reasonable to interpret the phase diagram of
a polycrystalline Y$_2$BaCuO$_5$ similarly to the phase diagram
that would be obtained for a single crystal with $\mathbf{H}$
parallel to the easy axis or to the easy plane (depending on the
nature of the anisotropy).

The extrapolation of the "high-field" linear behavior of $T_N(H)$
to zero field leads to a temperature $T_{N,0}$ smaller than
$T_N(H=0)$ by $\Delta=0.3$ K. As proposed for
BaNi$_{2}$V$_{2}$O$_{8}$, Sr$_2$CuO$_2$Cl$_2$, and Pr$_2$CuO$_4$
\cite{knafo07a} (see also Section \ref{similaritiesquasi2D}), we
speculate that, in Y$_2$BaCuO$_5$, $T_{N,0}$ is a virtual ordering
temperature, which would characterize the system in the limit of
no easy-axis anisotropy. In this picture, the increase of
$T_N(H=0)$ by $\Delta$ is a consequence of the intrinsic easy-axis
anisotropy. A linear extrapolation of the high-field variation of
$\Delta S_{N}(H)$ also leads to $\Delta S_{N,0}/\Delta
S_{mag}=(2.3\pm0.5)*10^{-3}$ at $H=0$ (cf. Fig. \ref{entropy}). As
well as $T_{N,0}$, we associate the extrapolated entropy change
$\Delta S_{N,0}/\Delta S_{mag}$ with the limit of no easy-axis
anisotropy. At zero magnetic field, the N\'eel temperature
$T_{N}(H=0)$ and the associated change of entropy $\Delta
S_{N}(H=0)$ have non-zero values, probably because of the
combination of the magnetic anisotropy (XY and Ising) and of a 3D
character of the magnetic exchange \cite{note_LRO}.

\subsection{Similarities between the phase diagram of Y$_2$BaCuO$_5$ and those of the quasi-2D BaNi$_{2}$V$_{2}$O$_{8}$, Sr$_2$CuO$_2$Cl$_2$, and
Pr$_2$CuO$_4$} \label{similaritiesquasi2D}

\begin{figure}[t]
    \centering
    \epsfig{file=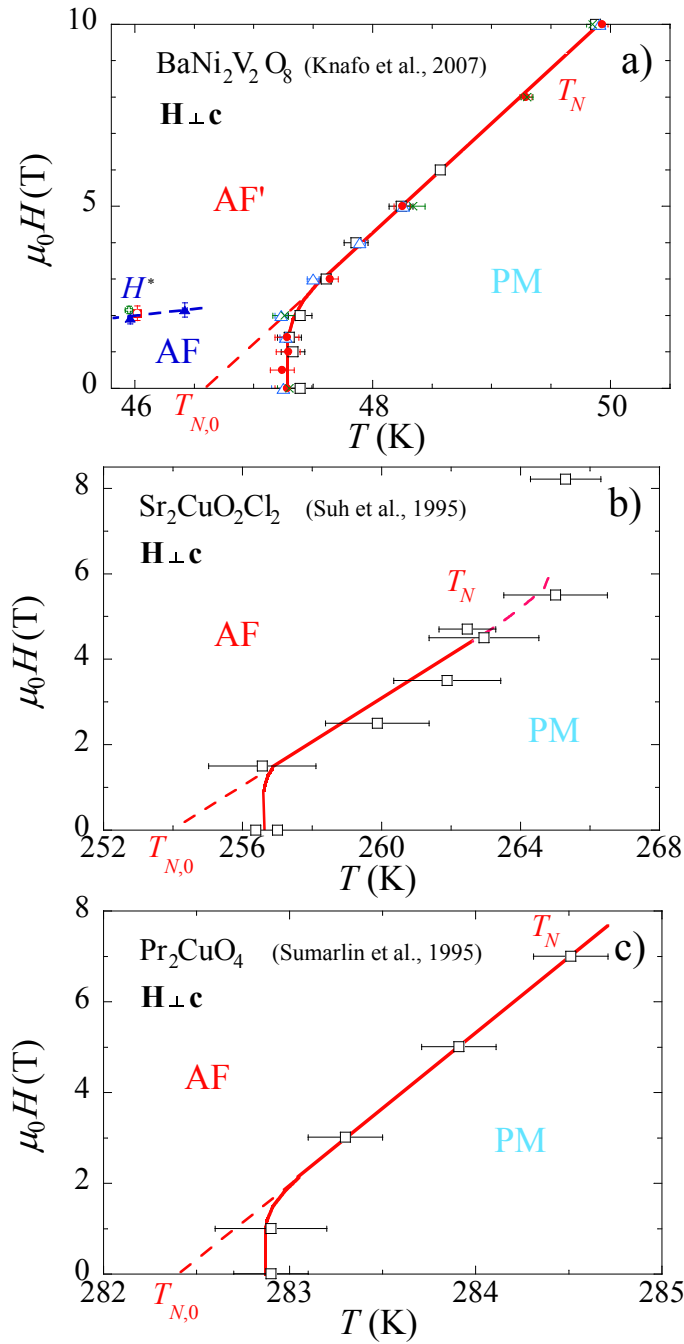,width=90mm}
    \caption{(color online) ($T,H$) phase diagrams of the quasi-2D magnetic systems (a) BaNi$_2$V$_2$O$_8$
    \cite{knafo07a}, b) Sr$_2$CuO$_2$Cl$_2$ \cite{suh95}, and c) Pr$_2$CuO$_4$
    \cite{sumarlin95}, with $\mathbf{H}\perp\mathbf{c}$. The data in (b) and (c) were
    scanned from Ref.\cite{suh95} and \cite{sumarlin95}.}
    \label{phase_diagram_quasi2D}
\end{figure}

\begin{table}[b] \caption{Characteristics of the magnetic properties
of Y$_2$BaCuO$_5$, BaNi$_2$V$_2$O$_8$, Sr$_2$CuO$_2$Cl$_2$, and
Pr$_2$CuO$_4$.}
\begin{center}
\label{table} \item[]\begin{tabular}{llcccc} \br
&& Y$_2$BaCuO$_5$ & BaNi$_2$V$_2$O$_8$ & Sr$_2$CuO$_2$Cl$_2$ & Pr$_2$CuO$_4$ \\
\mr
\multicolumn{2}{l}{$T_{max}$ (K) $^{*}$ $^{\natural}$} & 30 & 150 & n.d. & n.d. \\
\multicolumn{2}{l}{$T_{N}(H=0)$ (K)}& 15.6 & 47.4 & 256.5 & 282.8 \\
\multicolumn{2}{l}{$\Delta S_{N}(H=0)/\Delta S_{mag}$}& 4*10$^{-3}$ & 7*10$^{-4}$ & n.d. & n.d. \\
\;\;\;&\multicolumn{1}{|l}{$\mathbf{H}$ $^{\dag}$}& (polycrystal) & $\perp\mathbf{c}$ &  $\perp\mathbf{c}$ &  $\perp\mathbf{c}$ \\
\;\;\;&\multicolumn{1}{|l}{$\mu_0H^*$ (T) $^{\dag}$}& $\simeq2$ &  $\simeq1.5$ &  $\simeq1.5$ &  $\simeq2$ \\
\;\;\;&\multicolumn{1}{|l}{$T_{N,0}$ (K) $^{\dag}$} & 15.3 & 46.6 & 254  & 282.4  \\
\;\;\;&\multicolumn{1}{|l}{$a$ (K/T) $^{\dag}$} & 0.14 & 0.34 & 2  & 0.3  \\
\;\;\;&\multicolumn{1}{|l}{$\Delta S_{N,0}/\Delta S_{mag}$ $^{\dag}$} &  \;2.3*10$^{-3}$ & n.d. & n.d. & n.d. \\
\multicolumn{2}{l}{Symmetry}& orth. & hex. & tetr. & tetr. \\
\multicolumn{2}{l}{Ref.}& \cite{ong88}& \cite{knafo07a,rogado02} & \cite{suh95} & \cite{sumarlin95}\\
\br
\end{tabular}
\end{center}
\small
$^{*}$ : defined as the temperature of the maximum of $\chi(T)$\\
$^{\natural}$ : n.d. = non determined, hex. = hexagonal, orth. =
orthorhombic, tetr. = tetragonal\\
$^{\dag}$ : $T_N(H)=T_{N,0}+aH$ and $\Delta S_N(H)\simeq\Delta
S_{N,0}+bH$ for $H \geq H^*$
 \normalsize
\end{table}

As mentioned above, the $T$-$H$ phase diagram of polycrystalline
Y$_2$BaCuO$_5$, shown in Fig. \ref{phase_diagram}, has a striking
resemblance with the phase diagrams of single crystals of the
quasi-2D antiferromagnets BaNi$_{2}$V$_{2}$O$_{8}$
\cite{knafo07a}, Sr$_2$CuO$_2$Cl$_2$ \cite{suh95}, and
Pr$_2$CuO$_4$ \cite{sumarlin95}, which are shown in Fig.
\ref{phase_diagram_quasi2D} (a), (b), and (c), respectively, for
$H$ applied within the easy plane \cite{note_phase_diagrams}. In
these insulating systems, $T_N(H)$ is constant for $H \leq H^*$
and increases with $H$ for $H \geq H^*$. A linear increase of
$T_N(H)$ is unambiguously obtained in the high-field regime of
Y$_2$BaCuO$_5$ and BaNi$_{2}$V$_{2}$O$_{8}$ \cite{knafo07a} and is
compatible, within the experimental errors, with the phase
diagrams of Sr$_2$CuO$_2$Cl$_2$ \cite{suh95} and Pr$_2$CuO$_4$
\cite{sumarlin95}. In Table \ref{table}, $T_N(H=0)$, $T_{N,0}$,
$H^*$, and $a$ (from a fit by $T_N(H)=T_{N,0}+aH$ for $H>H^*$) are
given for each of these systems. For Y$_2$BaCuO$_5$ \cite{ong88}
and BaNi$_{2}$V$_{2}$O$_{8}$ \cite{rogado02}, the temperature
$T_{max}$ of the maximum of the magnetic susceptibility $\chi(T)$,
characteristic of the low-dimensional magnetic exchange, is also
given. The investigation of the magnetic properties of these two
systems is rather easy, since their full magnetic entropy is
contained below room temperature (see Section \ref{results} and
Ref. \cite{knafo07a}), as illustrated by the rather small values
of $T_N$ and $T_{max}$, which are of the order of several tens of
Kelvin. Oppositely, the magnetic properties of Sr$_2$CuO$_2$Cl$_2$
and Pr$_2$CuO$_4$, as well as those of the underdoped high-$T_C$
cuprates, are more difficult to investigate, being associated with
temperature scales $T_{max}>T_N\simeq300$ K \cite{notemaxsuscept}.

We interpret the enhancement of $T_N(H)$ with increasing magnetic
field in BaNi$_{2}$V$_{2}$O$_{8}$, Sr$_2$CuO$_2$Cl$_2$, and
Pr$_2$CuO$_4$, as well as the one observed in Y$_2$BaCuO$_5$,
using an effective field-induced anisotropy \cite{villain77}.
Contrary to the present work, which was made using a polycrystal
of Y$_2$BaCuO$_5$, the studies of BaNi$_{2}$V$_{2}$O$_{8}$,
Sr$_2$CuO$_2$Cl$_2$, and Pr$_2$CuO$_4$ were performed on single
crystals with $\mathbf{H}\perp\mathbf{c}$
\cite{knafo07a,suh95,sumarlin95}. While the magnetic properties of
Y$_2$BaCuO$_5$ cannot be easily related to its 3D and rather
complex crystal structure (Fig. \ref{structure}), the quasi-2D
magnetic exchange paths of BaNi$_{2}$V$_{2}$O$_{8}$,
Sr$_2$CuO$_2$Cl$_2$, and Pr$_2$CuO$_4$ are a direct consequence of
their layered crystallographic structure, and their intrinsic
anisotropy is controlled by the symmetry of their lattice. The
intrinsic in-plane anisotropy, hexagonal for
BaNi$_{2}$V$_{2}$O$_{8}$ \cite{rogado02} and tetragonal for
Sr$_2$CuO$_2$Cl$_2$ \cite{suh95} and Pr$_2$CuO$_4$
\cite{sumarlin95}, leads to magnetic domains at zero field where
the spins align (in the easy plane) along one of three equivalent
easy axes in BaNi$_{2}$V$_{2}$O$_{8}$ and along one of two
equivalent easy axes in Sr$_2$CuO$_2$Cl$_2$ and Pr$_2$CuO$_4$
\cite{suh95,sumarlin95,rogado02}. In Ref. \cite{suh95}, Suh et al.
proposed that the change of behavior of $T_N(H)$, which occurs at
$\mu_0H^*\simeq2$ T in Sr$_2$CuO$_2$Cl$_2$, is related to a
field-induced crossover from a regime controlled by the intrinsic
XY anisotropy to a regime controlled by the field-induced Ising
anisotropy. For Sr$_2$CuO$_2$Cl$_2$, but also for
BaNi$_{2}$V$_{2}$O$_{8}$ and Pr$_2$CuO$_4$, we propose that the
change of behavior of $T_N(H)$ at $H^*$ results in fact from a
crossover between a regime controlled by the intrinsic Ising-like
in-plane anisotropy (with two or three equivalent easy axes) to a
regime controlled by the field-induced Ising anisotropy (with one
easy axis) \cite{knafo07a}. Although the XY anisotropy plays an
important role in both regimes above and below $H^*$, we believe
that the crossover at $H^*$ is not directly related to the XY
anisotropy, as proposed in Ref. \cite{suh95}, but is due to the
small residual Ising-like anisotropy, which ultimately determines
the easy axis within the XY plane.

We furthermore speculate that, for Sr$_2$CuO$_2$Cl$_2$ and
Pr$_2$CuO$_4$, as well as for BaNi$_{2}$V$_{2}$O$_{8}$
\cite{knafo07a}, $T_{N,0}$ corresponds to a "virtual" N\'{e}el
temperature which would be achieved in a limit with no in-plane
anisotropy, being controlled by the combination of the 2D
exchange, the XY anisotropy, and the interlayer 3D exchange. Thus,
$T_{N,0}$ would give an upper limit of the
Berezinskii-Kosterlitz-Thouless temperature $T_{BKT}$,
characteristic of a pure 2D XY magnetic system \cite{note_LRO}.
Moreover, $T_{N,0}$ and $T_{BKT}$ would be equal if the interlayer
exchange would be negligible. This would also apply for
Y$_2$BaCuO$_5$ if one could prove, e.g., by neutron scattering,
that it is quasi-2D (thus not quasi-1D). Since a strictly
low-dimensional (1D or 2D) Heisenberg system corresponds to a
limit with no transition \cite{note_LRO}, thus with $\Delta
S_{N}\rightarrow0$, the fact that $\Delta S_{N}(H=0)/\Delta
S_{mag}$ is six times smaller in BaNi$_{2}$V$_{2}$O$_{8}$ than in
Y$_2$BaCuO$_5$ (Table \ref{table}) indicates a stronger
low-dimensional character in BaNi$_{2}$V$_{2}$O$_{8}$ than in
Y$_2$BaCuO$_5$. The non-zero values of $T_{N,0}$ and $\Delta
S_{N,0}$ in Y$_2$BaCuO$_5$ are the consequence of the XY
anisotropy and/or of the 3D exchange, in addition to the
low-dimensional exchange (1D or 2D) \cite{note_LRO}.

In Ref. \cite{bloembergen77} Bloembergen compared the
specific-heat data of several quasi-2D magnetic systems. He
discussed how the residual 3D intralayer exchange, as well as XY
and in-plane anisotropies, stabilize long-range ordering in these
systems, leading to an increase of $T_N$ and of the size of the
associated specific-heat anomaly. Our data, but also our
interpretation, are very similar to the ones of Bloembergen
\cite{bloembergen77} (cf. the heat-capacity data from Fig.
\ref{spec_heat_TN} of the present article and from Fig. 2 and 3 of
Ref. \cite{bloembergen77}). As an important new feature, our work
shows  that a magnetic field can be used to continuously tune the
magnetic anisotropy of low-dimensional systems.

\begin{figure}[t]
    \centering
    \epsfig{file=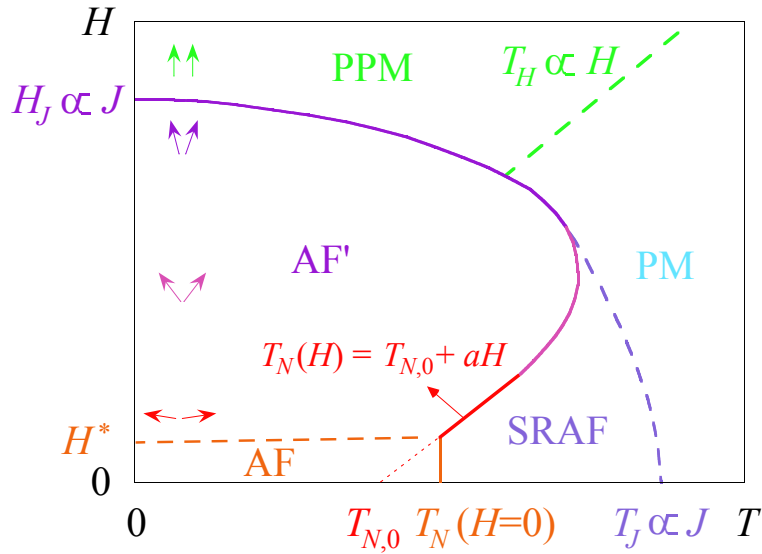,width=100mm}
    \caption{(color online) Phase diagram expected at high enough magnetic
    fields for Y$_2$BaCuO$_5$, BaNi$_{2}$V$_{2}$O$_{8}$, Sr$_2$CuO$_2$Cl$_2$,
    and Pr$_2$CuO$_4$ (AF = low-field antiferromagnetic phase,
    AF' = field-induced canted phase, PPM = high-field polarized
    paramagnetic phase, PM = high-temperature paramagnetic regime,
    and SRAF = low-dimensional short-range antiferromagnetic regime).}
    \label{phase_diagram_high_fields}
\end{figure}

In Fig. \ref{phase_diagram_high_fields}, we present a tentative
extension to very high magnetic fields of the phase diagrams of
Y$_2$BaCuO$_5$, BaNi$_{2}$V$_{2}$O$_{8}$, Sr$_2$CuO$_2$Cl$_2$, and
Pr$_2$CuO$_4$. In these systems, the low-dimensional (1D or 2D)
exchange $J$ is the dominant magnetic energy scale, and a
ferromagnetic polarized regime, where the spins are aligned
parallel to the magnetic field, has to be reached at magnetic
fields $H>H_J$, with $H_J\propto J$. This implies that the phase
line $T_N(H)$, which first increases linearly, as reported here,
will reach a maximum before decreasing down to zero at $H_J$, the
spins being in a canted antiferromagnetic state for $H^*<H<H_J$.
As evidenced by the broad maxima observed in the heat capacity and
magnetic susceptibility measurements, a crossover occurs at
zero-field at $T_J\propto J$, which corresponds to the onset of
low-dimensional short-range ordering and consists of
low-dimensional antiferromagnetic fluctuations. The application of
a magnetic field is expected to decrease $T_J$, and we speculate
that $T_J(H)$ will merge at very high fields with the transition
line $T_N(H)$ of the canted ordered state, since $T_N(H)$ will
cancel out at $H_J\propto J\propto T_J(H=0)$. Finally, a crossover
characteristic of the field-induced polarized state is expected to
occur at $T_H\propto H$. Pulsed magnetic fields should be used to
extend the phase diagram presented in Fig. \ref{phase_diagram} up
to much higher fields, in order to try to reach its polarized
state and to check the validity of the tentative phase diagram of
Fig. \ref{phase_diagram_high_fields}. This should enable one to
extract the different magnetic energy scales and lead to a better
understanding of the magnetic properties of Y$_2$BaCuO$_5$.

\section{Conclusion}

The $T$-$H$ phase diagram of the low-dimensional magnetic system
Y$_2$BaCuO$_5$ determined by heat-capacity measurements has
revealed striking resemblances with the phase diagrams of the
quasi-2D magnetic systems BaNi$_{2}$V$_{2}$O$_{8}$,
Sr$_2$CuO$_2$Cl$_2$, and Pr$_2$CuO$_4$. Although we do not know
the nature of the exchange (quasi-1D or quasi-2D) and of the
magnetic anisotropy in Y$_2$BaCuO$_5$, we interprete the increase
of $T_N(H)$ as resulting from a field-induced anisotropy. In this
scenario, at the lowest energy or corresponding field scales, the
Ising-like anisotropy is important. Further, our work permitted to
demonstrate that external magnetic fields can be used to
continuously tune an effective spin anisotropy. We observed a
field-induced transfer of magnetic entropy from the
low-dimensional high-temperature broad signal to the anomaly
associated with the 3D ordering at $T_N$, which is related to the
presence of an isosbestic point at $T_X\simeq20$ K.

The comparison of the magnetic properties of Y$_2$BaCuO$_5$,
BaNi$_{2}$V$_{2}$O$_{8}$, Sr$_2$CuO$_2$Cl$_2$, and Pr$_2$CuO$_4$
should be useful to refine theoretical models. New theoretical
developments are needed to understand these properties on a more
quantitative level, notably the linear increase of $T_N$ with $H$
at moderate magnetic fields. The theories, which already consider
the XY anisotropy and the different kinds of exchange interactions
(see for example Ref. \cite{cuccoli03}), should be extended to
include an Ising-like anisotropy term, which ultimately determines
the direction of the ordered spins and stabilizes long-range
ordering. Also, because of their small energy scales, the magnetic
properties of Y$_2$BaCuO$_5$ and BaNi$_{2}$V$_{2}$O$_{8}$ can be
accessed rather easily and their study may yield significant clues
for understanding the magnetic properties of the high-$T_C$
cuprates.

In the future, single crystals of Y$_2$BaCuO$_5$ should be studied
to determine the nature of the anisotropy and of the magnetic
exchange (e.g., from susceptibility and neutron scattering
measurements). Pulsed magnetic fields could also be used to extend
the phase diagram of Y$_2$BaCuO$_5$ to higher fields. Finally, it
would be interesting to try to decrease by doping the strength of
the exchange interactions, in order to enter into a paramagnetic
conducting phase and to study the properties of the related
metal-insulator crossover.

\section*{Acknowledgments}

We would like to thank K.-P. Bohnen and R. Heid for providing us
the ab-initio calculations of the phonon contribution to the
specific heat of YBa$_2$Cu$_3$O$_7$. This work was supported by
the Helmholtz-Gemeinschaft through the Virtual Institute of
Research on Quantum Phase Transitions and Project VH-NG-016.


\begin{thebibliography}{10}
\bibitem{noteentropy}
For each magnetic field, $(\Delta C_p^{mag}/T)_{max}$ corresponds
to the maximal value of $\Delta C^{mag}_p(T)/T$ and $\Delta S_{N}$
is obtained by integration of $\Delta C^{mag}_p(T)$ (Fig.
\ref{spec_heat_TN} (b)).
\bibitem{noteTN}
The shape of the antiferromagnetic anomaly (Fig.
\ref{spec_heat_TN} (b)) is characteristic of a second-order phase
transition broadened by some sample inhomogeneities. While a pure,
thus non broadened, second order transition leads to a step-like
anomaly at $T_N$ in the specific heat, the present case consists
in a broad transition where $T_N$ can be defined at the minimum of
slope of $C_p(T)$.
\bibitem{note_phase_diagrams} While the phase diagram of Y$_2$BaCuO$_5$ was obtained here from the
specific heat, those of BaNi$_{2}$V$_{2}$O$_{8}$,
Sr$_2$CuO$_2$Cl$_2$, and Pr$_2$CuO$_4$ were obtained from specific
heat and thermal expansion \cite{knafo07a}, magnetization and NMR
\cite{suh95}, and neutron scattering \cite{sumarlin95},
respectively.
\end{thebibliography}
\end{document}